\documentstyle[sprocl]{article}

\def\Journal#1#2#3#4{{#1} {\bf #2}, #3 (#4)}

\def\be{\begin{equation}}
\def\ee{\end{equation}}
\def\bea{\begin{eqnarray}}
\def\eea{\end{eqnarray}}

\begin{document}

\title{
DETAILED CHEMICAL ANALYSIS OF TWO GIANTS IN THE SGR DSPH 
}

\author{P. BONIFACIO, P. MOLARO}

\address{Osservatorio Astronomico di Trieste, Via G.B. Tiepolo 11, 
\\ 34131 - Trieste, Italy\\E-mail: bonifaci@ts.astro.it, molaro@ts.astro.it}


\maketitle\abstracts{ 
The 8m class telescopes allow for the first time to study stars of
external galaxies with the same resolution and S/N ratio which has been
so far used for Galactic stars. It is quite likely that this study will
shake some of our current beliefs.
In this poster we highlight some of the results which have been obtained
for two giants in the Sgr dSph thanks to the UVES spectrograph on the
ESO 8.2m Kueyen telescope. Further details on the observations and
data analysis may be found in Bonifacio et al \cite{b00}.
}

\section{Results}
The two stars turned out to be rather metal-rich with [Fe/H] $\approx -0.3$.
This was somewhat a surprise since the stars have roughly the same $V-I$
colour but differ in $V$  by 0.18 mag and were thought to
be representative of the spread in Sgr RGB which common--wisdom
has attributed to a spread in metallicity\cite{i95,sl95,mat95,i97,m98,bel} .
Given the steepness of the RGB in this colour range it may well be that
the two stars lie in fact on the same isochrone and their $V-I$ colours
are different but photometric errors of $\approx 0.04$ mag make them equal.
However there is another possible explanation whose consequences
are rather intriguing and cannot be easily discarded: the two stars
have a different distance. A difference of about 2 kpc  would be enough
to explain the difference in $V$ mag. This value is not unreasonable
since Ibata et al \cite{i97} estimate a line--of--sight depth of about
1.2 kpc.
From our spectra we may find evidence which further supports this
possibility: 
the Na I D lines of the brighter star,
\# 143, appear perfectly symmetric with no hint of the presence of interstellar
components at the radial velocity of Sgr, while the fainter
star, \# 139, shows a weak but definite asymmetry in both Na I D lines,
which may be interpreted as due to interstellar material associated with
Sgr.

The high metallicity found for both stars implies that
Sgr has undergone a high level of chemical processing; it also
suggests that a sizeable population of Sgr is this metal--rich.
The abundance pattern is interesting,
although most of the elements with $A\le 39$ are consistent
within error with the solar ratio, taken at face value they
are all slightly sub-solar. On the other hand Ba and more massive
elements show over-solar ratios. The pattern may be summarized as follows:
 Na and Sc are  over-deficient with respect to iron
                 by about  $0.4$ dex;
 heavy neutron--capture elements
                 Ba, La, Ce, Nd, Eu are over--abundant
                 with respect to iron by $0.3$ to $0.5$ dex;
Y is over-deficient with respect to iron by about 0.4 dex.

It is interesting to compare the abundances of these two Sgr giants
with those obtained by Shetrone et al \cite{sh} for three other
Local Group Galaxies (Draco, Ursa Minor and Sextans). 
One should also keep in mind that Sagittarius is considerably more massive
than the other three galaxies and this is, perhaps, the key to understand
its high--metallicity population. The [Mg/Fe] ratio seems to drop,
with increasing [Fe/H], faster in Local Group Galaxies than it
does in the Galactic Halo or disk and this could be the signature of a very
low or episodic star formation. It shall be very interesting to see
what is the [Mg/Fe] ratio of the metal--poor population of Sagittarius,
when its existence is confirmed.
Damped Lyman $\alpha$ systems, observed in absorption in the line--of--sight
of QSO's have metallicities in the range -2,-1 and show [$\alpha$/Fe]
ratios which are solar \cite{cen,mol00}.
Could Damped Lyman $\alpha$ systems be linked to dwarf galaxies ?
This possibility is appealing, although these galaxies are now
devoid of gas, they must have been relatively rich in gas when the stars
we observe now were formed.

The over-deficiency of Na seems a common occurrence
in Local Group galaxies, although there is
no definite trend and a large scatter.
On the other hand Y is always over-deficient with very few exceptions
and this over-deficiency is mirrored by an overabundance of heavy elements.


\section*{References}
\begin{thebibliography}{99}
\bibitem{b00}
Bonifacio P., et al \Journal{\em A\&A}{359}{66}{2000} 
\bibitem{i95}
Ibata R.A., Gilmore G., Irwin M.J. \Journal{\em MNRAS}{277}{781}{1995}
\bibitem{sl95}
Sarajedini A., Layden A.C. \Journal{\em AJ}{109}{1086}{1995}
\bibitem{mat95}
Mateo M., 
et al \Journal{\em AJ}{109}{588}{1995}
\bibitem{i97}
Ibata R.A., et al,
\Journal{\em AJ}{113}{634}{1997}
\bibitem{m98}
Marconi G., 
et al \Journal{\em A\&A}{330}{453}{1998}
\bibitem{bel}
Bellazzini, M., Ferraro, F.R., Buonanno R., \Journal{\em MNRAS}{307}{619}{1999}
\bibitem{sh}
Shetrone M.D., C\^ ot\' e P., Sargent W.L.W. 
{\em ApJ} in press
 astro-ph/0009505 (2000)
\bibitem{cen}
Centuri\`on M, Bonifacio P., Molaro P., Vladilo G.
\Journal{\em ApJ}{536}{540}{2000}
\bibitem{mol00}
Molaro P., et al \Journal{\em ApJ}{541}{54}{2000}



\end{thebibliography}
\end{document}